\documentclass[twocolumn,showpacs,preprintnumbers,amsmath,amssymb]{revtex4}
\usepackage{graphicx}
\usepackage{dcolumn}
\usepackage{bm}

\newcommand{\be}{\begin{equation}}
\newcommand{\ee}{\end{equation}}
\newcommand{\bea}{\begin{eqnarray}}
\newcommand{\eea}{\end{eqnarray}}

\begin{document}
\title{Towards large $r$ from $[p,q]$-inflation}
\author{Eran Palti and Timo Weigand}
\affiliation{%
Institut f\"ur Theoretische Physik, Ruprecht-Karls-Universit\"at Heidelberg,
             Heidelberg, Germany.
}
\begin{abstract}
The recent discovery of B-mode polarizations in the CMB by the BICEP2 collaboration motivates the study of large-field inflation models which can naturally lead to significant tensor-to-scalar ratios. A class of such models in string theory are axion monodromy models, where the shift symmetry of an axion is broken by some branes. In type IIB string theory such models so far utilized NS5 branes which lead to a linear potential with an induced tensor-to-scalar ratio of $r \sim 0.07$. In this short note we study a modification of the scenario to include $[p,q]$ 7-branes and show that this leads to an enhanced tensor-to-scalar ratio $r \sim 0.14$. Unlike 5-branes, 7-branes are in-principle compatible with supersymmetry, however we find that an implementation of the inflationary scenario requires an explicit breaking of supersymmetry by the 7-branes during inflation. This leads to similar challenges as in 5-brane models. We discuss the relation to high-scale supersymmetry breaking after inflation.

\end{abstract}
\maketitle

\section{Introduction}
\label{sec:intro}

An interesting class of inflationary models are those which involve large (super-Planckian) field ranges because they induce primordial tensor modes that can potentially be measured as B-mode polarization of the CMB \cite{Inflation}. The BICEP2 collaboration has recently reported the detection of such B-modes \cite{Ade:2014xna} which may be primordial in origin. Provided this interpretation corroborates in the future, large-field models are therefore of interest as a possible explanation of these results. A universal challenge in constructing effective theories of large-field models of inflation is to control Planck-scale suppressed operators, which are expected to arise in quantum gravity. One possibility is to assume an approximate shift symmetry of the inflaton, so-called natural inflation \cite{Freese:1990rb}, which protects against such operators. It remains however of significant interest to understand how to embed such models in quantum gravity, string theory being currently our best candidate, in order to explicitly understand and control the shift symmetry and its corrections. 

An interesting class of string inflation models which aim at doing this are axion monodromy inflation models \cite{Silverstein:2008sg,McAllister:2008hb,Flauger:2009ab} \footnote{See also \cite{Berg:2009tg} for a variation of the models, and \cite{Dimopoulos:2005ac, Grimm:2007hs} for multi-axion models with potentially large tensor modes.}. In the type IIB version of these models the inflaton is an axion field arising in the Ramond-Ramond (RR) sector of string theory, which has a very well protected shift symmetry. This symmetry is broken by non-perturbative effects, and in particular branes. The idea of \cite{McAllister:2008hb} is to include an NS5 - anti-NS5 brane pair which induces a linear potential for the axion. This means that its compact field range is unraveled to a long (formally infinite) range thereby allowing for super-Planckian excursions along a direction which enjoys significant protection due to an approximate shift symmetry. This could then lead to a realisation of a version of natural inflation in string theory, with a significant tensor-to-scalar ratio of $r \simeq 0.07$ \cite{McAllister:2008hb}. 

The shift symmetry protection of the axions is quite well understood in the absence of the NS5-branes, and so is the potential for the axion induced by the branes when they are treated as {\it probes} in the background. What is less understood is the effect of the backreaction of the NS5-branes on the geometry and in particular the resulting modification of the four-dimensional effective scalar potential that is used for inflation. The backreaction was studied in \cite{McAllister:2008hb,Flauger:2009ab,Conlon:2011qp} though it is fair to say that it remains an open problem of critical importance to realising this scenario in string theory. One of the key technical obstacles to achieving a better handle on the models is that the introduced branes break supersymmetry explicitly. This supersymmetry breaking also has phenomenological consequences since it is expected that the scale of supersymmetry breaking in the vacuum after inflation is correlated with the dimensionful parameters in the inflaton potential. The recent BICEP2 measurements would then imply a high scale $\sim 10^{15}$ GeV of supersymmetry breaking in these models \footnote{The connection between BICEP2 and the supersymmetry breaking scale has been investigated in \cite{Ibanez:2014zsa}.}.

In this short note we perform preliminary investigations of a modification of this scenario that preserves the key properties of natural inflation but which uses different sources to break the shift symmetry and to induce the potential for the axion. Instead of using an NS5  - anti-NS5 pair we consider $[p,q]$ 7-branes. We will show that such branes induce a potential which can have terms up to order $b^{2}$ in the inflaton $b$. Under the crucial assumption that all technical obstacles associated with the backreaction of these branes can be resolved, this would lead to a chaotic inflation type model with amplified tensor modes $r \simeq 0.14$, which brings it closer to the BICEP2 measurement of $r = 0.2 \pm 0.05$ \cite{Ade:2014xna}\footnote{Other realisations of chaotic inflation via axion monodromies include \cite{Kaloper:2008fb}.}. 

Apart from this cosmological observational merit of this construction we are interested in the interplay with supersymmetry breaking. Unlike 5-branes, 7-branes can in principle preserve the supersymmetry of the Type IIB background under consideration, or at least break it spontaneously rather than explicitly. The question of which configurations of $[p,q]$ 7-branes preserve supersymmetry is beautifully understood using the technology of F-theory \cite{Vafa:1996xn}. It is thus natural to investigate whether an improved control over such constructions could be reached and whether the possibility of low-scale supersymmetry breaking after inflation can be realised. We find that actually also in the case of 7-branes the required brane sources for large-field inflation must break supersymmetry explicitly during inflation thereby leaving the (2-derivative) supergravity framework. This therefore puts them on a similar footing in terms of microscopic control as the 5-brane models of \cite{McAllister:2008hb}. We will also comment on  the status of supersymmetry breaking at the end of inflation, which, when present, due to the difference in the potential now naturally occurs at the scale $\sim 10^{13}$ GeV.

At the technical level, we wish to emphasise that our investigations are at the probe-brane level and when considering backreaction very similar problems to those encountered in the NS5-brane models of \cite{McAllister:2008hb} arise in the context of $[p,q]$ 7-branes. We will make some comments on the backreaction effects and suggest possible solutions but will not present any concrete well controlled realisation of the model including backreaction effects. This will remain an open problem for future work.

\section{The Model}
\label{sec:model}

\subsection{D7-branes and B-fields}

We consider a type IIB orientifold compactification on a Calabi-Yau 3-fold $X_3$ with orientifold action $(-1)^{F_L} \Omega \sigma$, where the  holomorphic involution $\sigma$ acting on $X_3$ gives rise to O3- and O7-orientifold -planes.
Crucial for this paper is the decomposition of forms on $X_3$ into orientifold-even and orientifold-odd
combinations as $H^{p,q}(X_3) = H^{p,q}_+(X_3) \oplus H^{p,q}_-(X_3)$\cite{Grimm:2004uq}. In particular, expanding the $B_2$- and $C_2$-field along orientifold-odd 2-forms as 
 \bea
 C_2 = c^a \omega_a, \qquad  B_2 = b^a \omega_a 
 \eea
with $\omega_a$ a basis of $H^{1,1}_-(X_3)$ gives rise to a set of $h^{1,1}_-(X_3)$ axionic fields in the effective 4-dimensional theory. 
Note that in addition to these continuous moduli, expansion of $B_2$ and $C_2$ along a basis $\omega_{\alpha}$ of $H^{1,1}_+(X_3)$ describes discrete, torsional background degrees of freedom, which play no role for us and will be set to zero for simplicity in the sequel.
The axions organize into combinations $G^a = c^a - \tau b^a$ with $\tau = C_0 + \frac{i}{g_s}$ the axio-dilaton which form the bosonic components of an ${\cal N}=1$ supermultiplet. Details of the ${\cal N}=1$ effective supergravity action can be found in \cite{Grimm:2004uq}. It is these axionic fields which play the role of the inflaton in models of axion monodromy inflation \cite{McAllister:2008hb,Flauger:2009ab} and, prior to that, also in the axion inflation models of \cite{Dimopoulos:2005ac,Grimm:2007hs}.
The rationale behind this is that axionic fields enjoy a shift symmetry which helps keeping their scalar potential sufficiently flat for inflation to occur.
The shift symmetry is broken only by non-perturbative effects, certain fluxes and the coupling to suitable branes. The latter was argued in \cite{McAllister:2008hb} to enlarge the a priori periodic field range of the axion in such a way as to generate large field models of inflation.

In this article we will investigate under what condition the coupling of the axionic fields to 7-branes can create a large-field inflationary potential.
As will be analyzed below, perturbative D7-branes induce a potential for the $b^a$-moduli. Were it not for the fact, discussed in detail in \cite{Grimm:2007hs} and \cite{McAllister:2008hb,Flauger:2009ab}, that these $b^a$-moduli appear in the K\"ahler potential and thus suffer from an eta-problem, this potential could be used as the inflationary potential for the inflaton $b^a$. By contrast, the $c^a$-moduli do not suffer from such an eta-problem \cite{Grimm:2007hs},\cite{McAllister:2008hb,Flauger:2009ab}. In our context, this suggests considering not perturbative D7-branes, but more general $[p,q]$ 7-branes.
In this section we will ignore this eta-problem in order to describe the general setup in the more familiar perturbative framework with $b^a$ the would-be inflaton, postponing a discussion of $[p,q]$ 7-branes to the next section.

Consider therefore a spacetime-filling D7-brane with world-volume ${\cal M}_i = {\cal M}^{1,3} \times D_i$  for $D_i$ a holomorphic 4-cycle.
Its dynamics is described by the action
\bea
S &=& S_{\rm DBI} + S_{\rm CS}, \\
S_{\rm DBI} &=& - \frac{2\pi}{\ell_s^8} \int_{{\cal M}_i} d^8\zeta e^{-\phi} \sqrt{- {\rm det}(G + (F_i+\iota^*B_2))}, \nonumber \\
S_{\rm CS} &=& -\frac{2\pi}{\ell_s^8} \int_{{\cal M}_i} \sum_{p} \iota^* C_{2p} \, {\rm tr} e^{F_i + \iota^*B_2} \sqrt{\hat A(TD)/ \hat A(ND)}, \nonumber
\eea
where  $\ell_s = 2 \pi \sqrt{\alpha'}$ is the string length and $g_s = e^{\phi}$ is the string coupling. 
We are interested in a compactification with stacks of $N_i$ such D7-branes together with their orientifold images along the cycles $D_i'=\sigma D_i$.
To  sum up the contribution of all branes and image branes in the effective action,
it is oftentimes convenient to introduce the combinations $D_i^{\pm} = D_i \pm D_i'$ where the minus sign denotes orientation reversal of the 4-cycle. One then thinks in terms of  combined D7-brane - image pairs along the invariant 4-cycles      $D_i^+$ \cite{Jockers:2004yj}. Importantly, this 4-cycle contains both orientifold-even and orientifold-odd  2-cycles such that one can introduce a basis $\omega_{\hat \alpha}$ of $H^{1,1}_+(D_i^+)$ and  
$\omega_{\hat a}$ of  $H^{1,1}_-(D_i^+)$. 
Note that only a subset of these 2-forms arise by pullback of corresponding 2-forms from $X_3$ to the brane. 
It is into these pullback forms that we have to expand $\iota^*B_2$. Gauge flux on a brane-image brane pair $D_i^+$ is expanded in terms of the full basis  $\omega^{\hat a}$ of  $H^{1,1}_-(D_i^+)$. In particular, not all fluxes need to arise 'by pullback from $X_3$'.

Consistency of the configuration requires the cancellation of all D7-brane, D5-brane and D3-brane RR-tadpoles. We will assume these topological conditions to be satisfied in the sequel. 
In particular, as described in \cite{Blumenhagen:2008zz}, cancellation of the 7-brane tadpole $\sum_i N_i (D_i + D_i') = 8 D_{O7}$ implies that the continuous B-field moduli $b^a$ decouple from the D5- and the D3-tadpole cancellation condition.
This means that while a non-trivial VEV of any of the $b^a$ does induce an RR-D5 and an RR-D3-tadpole on a brane-image brane pair, the overall tadpoles induced by these fields are guaranteed to cancel automatically in any consistent setup,
\bea \label{D3D5}
&& \sum_i N_i  \int_{D^+_i} (b^a \iota^* \omega_a) \wedge \iota^* \omega_b  =0 \qquad  \forall \omega_b \in H^{1,1}_-(X_3), \nonumber \\
&& \sum_i N_i \int_{D^+_i} (b^a \iota^* \omega_a) \wedge   (b^a \iota^* \omega_a) =0.
\eea
The F-term supersymmetry condition requires $F^{0,2}=0=F^{2,0}$ and the D-term supersymmetry condition takes the form
\bea \label{Dterm}
\int_{D_i} J \wedge (F_i + b^a \iota^* \omega_a ) =0
\eea
with $J$ inside the K\"ahler cone of the divisor $D_i$.
The appearance of certain combinations of $b^a$-moduli inside the D-term is a consequence of the fact that their partners $c^a$ are eaten by means of a 'geometric' St\"uckelberg mechanism \cite{Jockers:2004yj,Plauschinn:2008yd,Grimm:2011tb} in the presence of 7-branes for which $D_i^-$ is homologically non-trivial (i.e. brane and image brane wrap homologically different 4-cycles). For later purposes note that if the 4-cycle is invariant under $\sigma$ as a whole (but not necessarily pointwise) such that  $D_i^-=0$, the condition (\ref{Dterm}) is identically satisfied, in agreement with the fact the gauge theory on a 7-brane along such a cycle does not give rise to any $U(1)$ factor; thus there is no D-term condition and no St\"uckelberg mechanism.

As shown in \cite{Blumenhagen:2008zz}, if (\ref{Dterm}) is satisfied inside the K\"ahler cone, then the flux and $B$-field induced D3-charge on $D_i$  is non-negative, $ - \int_{D_i}   (F_i + b^a \iota^* \omega_a)^2  \geq 0$ .
In the sequel we are interested, for simplicity, in configurations with supersymmetric gauge flux $\int_{D_i} J \wedge F_i = 0$. Then a non-trivial VEV of the moduli $b^a$ that satisfies the D-term supersymmetry condition for all 7-branes $D_i$ necessarily induces only non-negative D3-charge along every 7-brane, $- \int_{D_i}   \iota^*(b^a \omega_a)^2  \geq 0$. Combined with the second equation in (\ref{D3D5}) this implies that $\int_{D^+_i} (b^a \iota^* \omega_a) \wedge   (b^a \iota^* \omega_a) = 0$ for all $D_i$. Therefore, in order for a VEV of $b_a$ to induce some D3-brane charge in a 7-brane tadpole cancelling configuration there are two options: Either it must violate the D-term supersymmetry condition, or some of the branes must be anti-D7-branes. In this case due to the opposite sign in the tadpole cancellation condition  the conclusion that all  $\int_{D^+_i} (b^a \iota^* \omega_a) \wedge   (b^a \iota^* \omega_a) = 0$ can be avoided.  This will become important momentarily.

\subsection{Axion monodromy with D7-branes}

Let us now compute the general form of the  potential for the $b^a$ fields in presence of a specific 7-brane along the cycle $D_i$. 
This contribution will then have to be summed up over all 7-branes and their image branes in the configuration.  
The 7-brane metric $G$ factors into a four-dimensional piece $g_{4}$ and an internal component $g$. To compute the potential for $B$  we furthermore focus on the components of $F$ and $\iota^*B$ along the internal directions parallel to the 4-cycle wrapped by the 7-brane.
The expansion of the internal components of the DBI-action
then gives 
\bea
S \supset -\frac{2\pi}{\ell_s^8} \int d^4 x \,  e^{-\phi} \,  \sqrt{-g_{4}}  \,   \Gamma_i     
\eea
with $\Gamma_i  =  \int_{D_i}d^4 z \sqrt{g + (F_i + \iota^*B)}$.
Following e.g. \cite{Haack:2006cy} and references therein,  we can use holomorphicity of the 4-cycle $D_i$  to write
\bea
\Gamma_i &=&  \sqrt{  ( {\rm Re}Z_i)^2 + ({\rm Im }Z_i)^2 }, \\
 {\rm Re}Z_i &=&   \frac{1}{2} \Big(  \int_{D_i} J \wedge J -  \int_{D_i}  (F_i + \iota^*B) \wedge  (F_i + \iota^*B)\Big), \nonumber \\
 {\rm Im}Z_i &=&  \int_{D_i}  J\wedge  (F_i + \iota^*B) \nonumber
\eea 
with $J$ the K\"ahler form of $X_3$ pulled back to $D_i$.

The properties of the effective action obtained by summing over the contributions from all branes depends crucially on whether ${\rm Re} Z_i >0$ for all $D_i$ or not. 
Consider first the situation where ${\rm Re} Z_i >0$ for $D_i$.
In the limit $|{\rm Im Z}_i| \ll |{\rm Re}Z_i|$  the effective action  can be brought in the form of a standard 4-dimensional ${\cal N}=1$ supergravity action by expanding 
$\Gamma_i = {\rm Re} Z_i + \frac{1}{2} {\rm Im}(Z_i)^2/ {\rm Re} Z_i + \ldots$ and interpreting the second term as a four-dimensional D-term potential. For $F_i=0$, the sum over ${\rm Re} Z_i$ for all branes and their image cancels by means of tadpole cancellation. In particular this is true for the sum over the induced D3-brane charges from $\int_{D_i} \iota^* B \wedge \iota^* B$. For $F_i\neq 0$ the remaining flux dependent terms are due to the gauge-flux induced superpotential     \cite{Jockers:2004yj}.
Recall furthermore that ${\rm Re} Z_i >0$ corresponds to the real part of the gauge kinetic function. In this limit ${\rm Im Z}_i \neq 0$ describes spontaneous D-term supersymmetry breaking. 
By contrast, if ${\rm Re} Z_i  < 0$ for at least some $D_i$, i.e. $\int_{D_i} \iota^* B \wedge \iota^* B > \int_{D_i} J \wedge J$, the large $B$-field VEV  breaks supersymmetry the hard way. 
Due to the appearance of the modulus in $\Gamma_i = |{\rm Re} Z_i| + \frac{1}{2} {\rm Im}(Z_i)^2/ |{\rm Re} Z_i| + \ldots$ the cancellation of the terms quadratic in $B$ is spoiled. However, the remaining terms cannot be written in terms of a 4-dimensional D-term or superpotential - a superpotential would have to be a holomorphic function of $G_2 = C_2 - \tau B_2$, but no potential for $C_2$ is induced. 

The 4-dimensional potential is extracted by expanding 
$\iota^*B = b^a \iota^*\omega_a$.
In general several linear combinations of axions $b^a$ can act as potential inflaton fields. In the sequel we consider for simplicity a situation with only a single such axion and choose the basis $\omega^a$ such that the inflating axion is $b^1$ and expand $B  =b^1 \, \omega_1 \equiv b \,   \omega$.
We are interested in the large-field regime of big field VEV $b$. Depending on the specific choice of background gauge fluxes $F_i$ and on the properties of the 2-form $\omega$ in $B_2 = b \omega$ we arrive at different shapes of the inflationary potential.

First consider the case where $\int_{D_i} (b  \iota^* \omega) \wedge (b \, \iota^*\omega) =0$ for all 7-branes, i.e. $b$ induces no D3 or anti-D3 charge.
A potential for $b$ is induced if supersymmetry is broken by the D-terms $\int_{D_i} J \wedge b \iota^*\omega \neq 0$.
In the limit of large $b$-field the potential takes the form
\bea \label{potlin}
V \simeq \frac{2 \pi}{\ell_s^4} \, e^{-\phi} \, v_1 \, b, \qquad \ell_s^4 v_1 = \sum_i N_i \Big| \int_{D_i^+} J \wedge \iota^*\omega \Big|.
\eea
This is the linear potential studied in \cite{McAllister:2008hb,Flauger:2009ab} for axion monodromy with D5 and anti-D5-branes. Note that the above realisation via 7-branes does not require the introduction of anti-branes. While for large $b$ during inflation we leave the supergravity regime $|{\rm Im}Z_i| \ll |{\rm Re} Z_i|$, the setup is smoothly connected to the supergravity regime as $b$ decreases. 
Depending on the specific choice of gauge flux along the 7-branes, the linear potential (\ref{potlin}) receives subleading corrections from lower powers $b^q$, which will lead to slight modifications of the inflationary dynamics.


A qualitatively very different regime corresponds to the situation where $\int_{D_i} (b  \iota^* \omega) \wedge (b \iota^*\omega) \neq 0$. 
For large $b$ the potential is now dominated by a quadratic term 
\bea \label{potquad}
V \simeq \frac{2 \pi}{\ell_s^4} \, e^{-\phi} \, v_2 \, b^2, \qquad  \ell_s^4 v_2= \sum_i N_i  \Big| \int_{D_i^+}  \iota^*\omega \wedge \iota^*\omega \Big|,
\eea
again modulo subleading corrections. This form of the potential cannot be achieved with (Abelian) 5-branes alone. 

There are two conceivable configurations that could give rise to such a scenario, both of which leave the framework of 4-dimensional spontaneously broken supergravity:
If all branes are D7-branes and no anti-D7-branes are present, then $v_2\neq 0$ requires that for some  D7-branes ${\rm Re}Z_i < 0$ for large $ b$ as analysed above.
As $b$ decreases, these branes undergo a dynamical transition back to ${\rm Re}Z_i > 0$ and supersymmetry is restored. 
The second possible configuration includes a setup of tadpole cancelling D7 and anti-D7-branes to begin with. Now a non-trivial net quadratic potential is compatible with RR tadpole cancellation and ${\rm Re}Z_i >0$ for all (anti-)branes at the cost of introducing anti-D7-branes. This would be the direct analogue of the configuration of D5 and anti-D5-branes envisaged in \cite{McAllister:2008hb,Flauger:2009ab}. 
As an important special case consider a configuration with D7 and anti-D7-branes, each on an invariant 4-cycle $D_i$ (not all pointwise invariant): As remarked below (\ref{Dterm}), no D-term arises, i.e. $\int_{D_i} J \wedge b \, \iota^*\omega \equiv 0$, 
and there is no St\"uckelberg coupling of the (anti-)branes to $C_2$. Nonetheless in the large field limit a quadratic potential is induced compatible with tadpole cancellation.

To summarize this section, ignoring the issue of an eta-problem for $b^a$, D7-branes induce either a linear potential (\ref{potlin}) or a quadratic potential (\ref{potquad}) for the $b^a$-moduli in the large field regime. 
In all these cases the system breaks supersymmetry hard during inflation:
in the case of  (\ref{potlin})  it leaves the supergravity regime during inflation by violating the relation   $|{\rm Im}Z_i| \ll |{\rm Re} Z_i|$  and  in the case of (\ref{potquad}), possibly in addition, by including anti-D7-branes or  violating the condition ${\rm Re}Z_i >0$.
If the system is such that $\int_{D_i} J \wedge b \iota^*\omega =0$ during inflation, then  anti-branes must necessarily be present and therefore supersymmetry is also broken at the end of inflation (provided these do not annihilate after inflation).

\section{$[p,q]$ 7-branes}
\label{sec:pq7}

So far we have studied D7-branes which couple to $B_2$ in the DBI action. However the associated four-dimensional fields $b^a$ do not enjoy a well-protected shift symmetry since they appear explicitly in the K\"ahler potential \cite{Grimm:2007hs}, \cite{McAllister:2008hb,Flauger:2009ab}
 \footnote{Note that at the level of the scalar potential they do enjoy a shift symmetry when combined with a shift in the K\"ahler moduli, but this is part of the no-scale structure and is broken by any effects which break no-scale.}. It is therefore more desirable to use the RR axions $c^a$ as the inflaton, which means that we require branes that generate an analogous structure to that studied in the previous section on D7-branes but for the $c^a$. Indeed, in type IIB string theory  the DBI action of $[p,q]$ 7-branes couples to the linear combination $A_2= p B_2 + q C_2$. A D7-brane corresponds to a $[1,0]$ 7-brane. 
 The DBI action for a general $[p,q]$ 7-brane is given in \cite{Bergshoeff:2006gs}.

Around each $[p,q]$ 7-brane the axio-dilaton undergoes a monodromy encoded in the $SL(2, \mathbb Z)$ matrix
\bea
M_{[p,q]} = \begin{pmatrix}   
1 -pq & p^2 \\
- q^2 & 1 + pq
\end{pmatrix}.
\eea
 Then for  a consistent set of $[p,q]$ 7-branes the total monodromy vanishes, which is the analogous statement to tadpole cancellation. Just like D7-branes, a set of $[p,q]$ 7-branes may also preserve supersymmetry depending on the background in which they are placed. 

Perhaps the simplest case for a candidate $[p,q]$ 7-brane that generates a potential for the RR axions $c^a$ is a $[0,1]$ 7-brane, which is the S-dual of a D7-brane so we may call it an NS7-brane. The DBI action for this brane in string frame is \cite{Bergshoeff:2006gs}
\bea
S = - \frac{2 \pi}{\ell_s^8}   \int_{{\cal M}_i} e^{-\phi} |\tau|^2 \sqrt{{\rm det} \Big(G +  |\tau|^{-1} (F_{NS} + \iota^*C_2 ) \Big)} \nonumber
\eea
with $\tau = C_0 + \frac{i}{g_s}$ and $F_{NS}$ the field strength associated with the vector-field on the [0,1] 7-brane.
In particular, a quadratic potential for $c$ in $C_2 = c \,  \omega$ is induced under the same conditions as before and given precisely by $(\ref{potquad})$ with $c$ instead of $b$. 

Analogously to \cite{McAllister:2008hb,Flauger:2009ab} in the case of 5-branes, one could then consider placing a configuration of an NS7 - anti-NS7-branes into the usual Type IIB background such that a quadratic potential for $c$ is induced, thereby avoiding the eta-problem for $b$. This of course \emph{assumes} that the backreaction of the NS7 - anti-NS7-brane pair on the IIB geometry is under control, in particular that corrections to the shift symmetry for $c$ are under control.
What complicates a full analysis of this question, among other things, is the fact that by construction we are forced beyond the regime of supergravity due to the hard breaking of supersymmetry as discussed in the previous section in the context of a D7-brane.

As in the analogous situation for D7-branes there is an interplay between the NS7-branes and the background axion fields, some of which may get eaten to give a mass to the world-volume gauge fields. 
St\"uckelberg couplings arise, for NS7-branes, by dimensional reduction of  the coupling $\int F_{NS} \wedge B_6$ in the Chern-Simons action, with $B_6$ the dual to the 2-form $B_2$. 
The appearance of St\"uckelberg couplings for a linear combination of $b^a$ in the 4-dimensional effective action requires that this combination have a 4-dimensional shift symmetry which is gauged by the coupling.
On the other hand, if the structure of the IIB effective action is to be unchanged by the NS7-branes, we know that this shift symmetry is not realised, which is precisely the origin of the eta-problem.
A minimal requirement for consistency of the setup is thus the absence of an effective St\"uckelberg coupling between the NS7-brane vectors and any linear combination of $b^a$. 
This means that the vector field on the NS7-brane corresponding to the 'diagonal' $U(1)$ must be absent. For D7-branes this is guaranteed for branes along invariant 4-cycles, in which case a stack of 7-branes gives rise to $SO(2N)$ or $Sp(2N)$ gauge groups. More generally, a St\"uckelberg coupling is avoided for brane configurations with simple gauge groups, which in the case of general $[p,q]$ 7-branes can include e.g. also exceptional gauge groups.

In these situations at least the obvious inconsistency between a combination of the $[p,q]$ 7-brane system and the IIB background with shift symmetry only for $c^a$ is circumvented.
Note that in absence of a gauging of axions also the associated D-term, which now would be proportional to $\int J \wedge \iota^*C_2$, is absent. This reasoning excludes the possibility of a linear inflationary potential (\ref{potlin}) for $c^a$. This is an important difference to the axion monodromy scenario \cite{McAllister:2008hb,Flauger:2009ab}
 based on 5-branes, where no St\"uckelberg gauging of the axions from $B_2$ or $C_2$ is induced by the branes.
 
 For D7-branes we established that a quadratic potential in absence of a D-term requires the presence of anti-D7-branes such that supersymmetry is broken hard even at the end of inflation.
 Verfiying a corresponding statement in the presence of general $[p,q]$ 7-branes is more involved and left for future work. Thus we can currently not exclude a scenario where after relaxation of the $c$-field supersymmetry is restored with $[p,q]$ 7-branes. This would be compatible with low-energy supersymmetry.

\section{Signatures for Inflation}

To analyze the inflationary dynamics we need to work in terms of canonically normalised fields as discussed in the present context in \cite{Grimm:2007hs,McAllister:2008hb,Flauger:2009ab}.
We present the analysis for the axionic $c^a$ moduli coupling to the $[p,q]$ 7-branes taking the role of the inflaton.
Dimensional reduction of the 10-dimensional kinetic terms gives
\bea
S_{\rm kin} &=&  - \frac{1}{2}  \int_{{\cal M}^{1,9}} \frac{2 \pi}{\ell_s^8 }  d C_2 \wedge \ast d C_2 \nonumber \\
&=& - \frac{1}{2} \int_{{\cal M}^{1,3}}  d^4 x \sqrt{-g_{4}} \partial c^a \partial c^b  \gamma_{ab} 
\eea
with
$\gamma_{ab} = \frac{2 \pi}{\ell_s^8}  \frac{1}{3} \int_{X_3} \omega_a \wedge \ast \omega_b$.
For an isotropic 3-fold $X_3$ of volume ${\cal V} \ell_s^6$ we can approximate \cite{Svrcek:2006yi}
\bea
 \int_{X_3} \omega_a \wedge \ast \omega_b = \ell_s^6 \,  {\cal V}^{1/3}.
\eea
In the simplified situation with a single axion $c$ this implies that the canonically normalised inflaton field $\Phi$ is related to $c$ as \cite{McAllister:2008hb}
\bea
c^2 = \frac{6}{g_s^2 M^2_{\rm Pl}} {\cal V}^{2/3} \Phi^2
\eea
with $M^2_{\rm Pl} = \frac{4 \pi}{\ell_s^2} e^{-2 \phi} {\cal V}$.   
Ignoring subleading corrections and focussing on the only possibility of a quadratic potential (\ref{potquad}) with $b$ replaced by $c$, the inflaton effective action in the large field regime takes the form of chaotic inflation \cite{Inflation}, 
\bea
S = \int d^4 x \sqrt{-g_4} \Big(  - \frac{1}{2} (\partial \Phi)^2 - \frac{1}{2} m^2 \Phi^2 \Big)
\eea
with 
\bea
m^2 = 3 g_s \frac{M_{\rm Pl}^2}{2 \pi} \frac{v_2}{{\cal V}^{4/3}} e^{-A}  .
\eea
Here we have included a general warp factor $e^{-A}$ as in \cite{McAllister:2008hb,Flauger:2009ab} which would arise for the 7-branes localised in a strongly warped region.

Let us briefly collect the well-known result of the standard analysis of the cosmological observables for this class of models:  The slow-roll parameters take the form
\bea
\epsilon = \frac{M_{\rm Pl}^2}{2} \Big(\frac{V'}{V}\Big)^2 ={M_{\rm Pl}^2} \frac{2}{\Phi^2}  , \quad \eta = M^2_{\rm Pl} \frac{V''}{V} = {M_{\rm Pl}^2} \frac{2}{\Phi^2} \nonumber
\eea
such that slow-roll occurs for large field VEVs $\Phi \gg M_{\rm Pl}$.
For the number e-folds $N =\int_{\Phi_{\rm end}}^{\Phi_0} \frac{d \Phi}{\sqrt{2 \epsilon}}$ to be $N=60$, the excursion of the inflaton field must therefore be such that $\Phi_0 \simeq 15.5 M_{\rm Pl}$. The resulting value of the tensor-to-scalar ratio is
\bea
r = 16 \epsilon|_{\Phi = \Phi_0} \simeq 0.14
\eea
with spectral index $n_s = (1 + 2 \eta - 6 \eta) |_{\Phi= \Phi_0} \simeq 0.96$ and $n_t = - 2 \epsilon |_{\Phi= \Phi_0} = -0.018$.
The value of the inflaton mass $m$ is fixed by the requirement of reproducing the observed scalar perturbations
\bea
\Delta^2_R |_{\phi = \phi_0} = \frac{1}{{12 \pi^2}} { \frac{V^3}{M_{\rm Pl.}^6 V'^2} }|_{\Phi = \Phi_0}
\eea
to be $m = 10^{-6} M_{\rm Pl}$.
Note in particular that the value of $r$ is twice as big as in the case of a linear inflationary potential, corresponding to the original axion monodromy model with 5-branes  \cite{McAllister:2008hb,Flauger:2009ab}. A different route towards realising chaotic inflation from axion monodromies has been investigated in \cite{Kaloper:2008fb}.

\section{Discussion}
\label{sec:disc}

In this short note we have performed a preliminary investigation of a large-field inflation scenario in type IIB string theory which generalises the axion monodromy inflation models in \cite{McAllister:2008hb}. In the scenario we have studied $[p,q]$ 7-branes break the shift symmetry of closed-string axion fields thereby inducing, to leading order, a quadratic potential for them. The potential is similar to chaotic inflation with resulting inflationary parameters $n_s \simeq 0.96$ and $r \simeq 0.14$. These are in decent agreement with BICEP2 results \cite{Ade:2014xna}. 

These parameters will receive corrections from sub-leading sources, for example world-volume fluxes, the particular combination of $[p,q]$ 7-branes, non-perturbative corrections to the K\"ahler potential \cite{Flauger:2009ab}, and possible flattening effects \cite{Dong:2010in}. A key find is that although, unlike the 5-branes of \cite{McAllister:2008hb}, 7-branes are in-principle compatible with supersymmetry, the appropriate inflationary potential  can only be induced by explicitly supersymmetry breaking 7-branes. While we cannot at present exclude the possibility that supersymmetry is restored at the end of inflation for general $[p,q]$ 7-branes, the best understood configurations we could find exhibit fairly high-scale supersymmetry breaking at the end of inflation \footnote{Note that since the brane tension required is $m \sim 10^{-6} M_{\rm Pl}$, such a scale of supersymmetry breaking when gravity mediated would lead to a gravitino mass estimate of $m_{3/2} \sim 10^{-12} M_{\rm Pl}$ which is not so high.}.

The construction outlined is far from an actual implementation in a well controlled string theory setting. Let us collect just a few of the challenges that must be overcome in constructing such a string theory vacuum. The most important issue is understanding the backreaction of the branes that break the shift symmetry on the geometry, and ensuring that this does not ruin the flatness of the potential or destabilises the whole vacuum. In the model studied in this note we find that 7-brane sources which explicitly break supersymmetry must be included. Since 7-branes backreact logarithmically this implies a serious issue of control. A similar backreaction problem for 5-branes was studied in \cite{McAllister:2008hb,Flauger:2009ab,Conlon:2011qp}. It was suggested in \cite{McAllister:2008hb,Flauger:2009ab} that warping effects can help to suppress the backreaction sufficiently. A possible solution would be to put the brane -  anti-brane pair in a bifurcated warped throat so that it only sources a dipole backreaction \cite{Flauger:2009ab}.
In \cite{Conlon:2011qp} it was further suggested though that due to tadpole cancellation between the NS5 and anti-NS5 there are also some 5-brane logarithmic effects that can leave the throat and therefore are not warp-factor suppressed either. Therefore in this sense the backreaction problems may be similar and must be understood further. 

Another key problem is that if the branes modify the 4-cycle volumes associated to the K\"ahler moduli significantly, then a KKLT \cite{Kachru:2003aw} type moduli stabilisation scenario using non-perturbative effects in the superpotential would lead to an eta-problem, thereby destroying the flatness of the potential \cite{McAllister:2008hb,Flauger:2009ab,Grimm:2007hs}. In the model we have studied this is an important problem particularly for the cycles associated to the 7-branes. One possible way to avoid this might be to fix the associated cycle perturbatively through corrections to the K\"ahler potential.

Finally, as was discussed already in \cite{Grimm:2007hs,McAllister:2008hb}, one must be careful to forbid the presence of non-perturbative corrections to the superpotential which involve the inflationary axions directly as these would induce an eta-problem. Such terms would come from fluxed E3-instantons, for example as studied in \cite{Grimm:2011dj}. On a positive note, the inflationary mechanism only requires that one such axion does not appear in the superpotential. On the other hand, one must sum over all such instantons in the superpotential with all possible world-volume fluxes. Whether it is possible to leave a protected axion seems a rather difficult and very model-dependent question.

We emphasise again that these are just some comments on the problems and in the absence of explicit computations it is difficult to make concrete statements either way regarding the backreaction, and more generally on a possible implementation in string theory of the scenario sketched in this note.

It is worth noting the following connection: the $[p,q]$ 7-branes used here are also the natural objects that appear when constructing Grand Unified Models (GUTs) in F-theory (see \cite{Weigand:2010wm} for reviews and references). It is an interesting coincidence that D7-branes alone are not sufficient for certain phenomenological requirements, in particular an order one top quark Yukawa coupling, and that one is forced to use the more general $[p,q]$ 7-branes. Further, the class of axions playing the role of the inflaton here can also appear in the gauge kinetic function of the visible sector 7-branes. Their presence was recently explored in \cite{Mayrhofer:2013ara} within a type IIB context. There it was shown that there is a rich interplay between such axions, certain properties of hypercharge flux restriction to matter curves, and resulting effects on gauge coupling unification. More generally it would be very interesting to explore such possible connections between GUT model building in F-theory and the proposed inflationary mechanism - particularly given the proximity of the inflationary energy scale to the GUT scale.

Finally, we would like to point out a related mechanism for coupling the axions to braneworld volumes. The 2-form fields leading to the axions 'canonically' couple to spacetime-filling 5-branes, and in this note we have studied their coupling to higher-dimensional branes. However it is also possible for them to couple to lower-dimensional branes, for example spacetime-filling D3-branes. This is because a stack of multiple coincident such branes has a coupling on its world-volume to all higher-dimensional fields as well \cite{Myers:1999ps}. The DBI term in the action for a $Dp$-brane contains the factor
\be
S_{DBI} \sim \sqrt{ \mathrm{det\;}Q^i_j} \;,
\ee
where $Q^i_j = \delta^i_j + i \lambda \left[\phi^i,\phi^k\right] \left( G_{kj} + i^*B_{kj} \right)$. Here the indices $i,j$ are orthogonal to the world-volume, and the $\phi^i$ are world-volume fields. Therefore even a D3-brane contains terms cubic in the NS axions $b^a$. The S-dual configuration would have such a coupling to the RR axions $c^a$. However, the problem is that for the potential to be non-vanishing the world-volume fields must develop a (non-commutative) vacuum expectation value. According to the Myers effect \cite{Myers:1999ps}, this happens  in the background of some external flux. However, what is harder to ensure is that there is no destabilisation to zero VEV for the world-volume fields when the axion VEV is large. This is similar to destabilisation of the closed-string moduli, but here warping effects are unlikely to help because the world-volume fields are also necessarily localised precisely where the energy density due to the axion VEV is localised. Nonetheless, inducing such potentials from non-Abelian branes remains an attractive possibility if a mechanism to fix the world-volume fields can be found. Indeed it is interesting to note that a non-Abelian stack of branes in IIB of any dimension will have a potential which is cubic for large axion VEVs, and that cubic potentials lead to an estimate of $r \sim 0.21$. \\

{\noindent \bf Acknowledgements}

We thank Arthur Hebecker, Sebastian Kraus  and Stefan Sjoers for useful and interesting discussions. EP is supported by the Heidelberg Graduate School for Fundamental Physics. This work was partially supported by the DFG under TR33 'The Dark Universe'.

\end{document}